# Broadband complementary vibrational spectroscopy with cascaded intra-pulse difference frequency generation

Kazuki Hashimoto[1], Venkata Ramaiah Badarla[1], Takayuki Imamura[2], and Takuro Ideguchi[*1,2]

[1] Institute for Photon Science and Technology, The University of Tokyo, Tokyo 113-0033, Japan

[2] Department of Physics, The University of Tokyo, Tokyo 113-0033, Japan

*ideguchi@ipst.s.u-tokyo.ac.jp

## Abstract

One of the essential goals of molecular spectroscopy is to measure all fundamental molecular vibrations simultaneously. To this end, one needs to measure broadband infrared (IR) absorption and Raman scattering spectra, which provide complementary vibrational information. A recently demonstrated technique called complementary vibrational spectroscopy (CVS) enables simultaneous measurements of IR and Raman spectra with a single device based on a single laser source. However, the spectral coverage was limited to ~1000 $cm^{-1}$, which partially covers the spectral regions of the fundamental vibrations. In this work, we demonstrate a simple method to expand the spectral bandwidth of the CVS with a cascaded intra-pulse difference-frequency generation (IDFG). Using the system, we measure broadband CVS spectra of organic liquids spanning over 2000 $cm^{-1}$, more than double the previous study.

Vibrational spectroscopy is an indispensable method to measure the functions and structures of molecular substances in a label-free manner. The molecular vibrational modes, categorized into IR- or Raman-active [1], are broadly lying in the fundamental vibrational wavenumber region (400-4000 $cm^{-1}$), and measuring all the fundamental vibrational modes is one of the ultimate goals of molecular spectroscopy. The comprehensive vibrational information allows one to investigate complex molecular phenomena and therefore benefit many applications, such as cancer diagnosis in biomedicine [2], complex catalytic analysis in chemistry [3], and bacterial analysis in food science [4], to name a few. Recent progress in advanced information technologies such as machine learning or deep learning strengthens the importance of multimodal and broadband spectroscopic measurements, which generate a large amount of data to be processed.

The full vibrational spectra may be obtained with a primitive combination of conventional techniques, e.g., a Fourier-transform infrared spectrometer (FTIR) and a grating-based spontaneous Raman spectrometer, which can individually measure broadband spectra (~3000 $cm^{-1}$). However, this approach suffers from instrumental complexity by combining the two different spectroscopy systems. Also, the different measurement modalities cause difficulty in accurate comparative analysis between the IR and Raman spectra due to the mismatch of the calibration methods. In addition, the inherently weak signal of the spontaneous Raman scattering severely limits the spectral acquisition rate, resulting in the bottleneck in measurement speed.

Recently, dual-modal spectrometers measuring IR and coherent Raman spectra in a single device with an ultrashort pulsed laser have been developed [5-7]. The spectrometers can measure IR spectra with mid-infrared (MIR) pulses generated via nonlinear wavelength-conversion processes from near-infrared (NIR) pulses, while Raman spectra with NIR pulses via coherent Raman scattering. The nonlinear optical effects provide advantages such as a brighter coherent MIR source and a faster measurement capability of coherent Raman spectra. However, these systems so far have a limited spectral bandwidth up to ~1000 $cm^{-1}$. Among the dual-modal spectrometers, complementary vibrational spectroscopy (CVS) [7], a combination of FTIR and Fourier-transform coherent anti-Stokes Raman scattering spectroscopy (FT-CARS) [8], has shown great potential to measure broader spectra because of the multiplex capability of Fourier-transform spectroscopy (FTS). However, in the previous demonstration with a 10-fs Ti:Sapphire (Ti:S) model-locked laser, the bandwidth of the IR spectrum was about 1000 $cm^{-1}$ (800-1800 $cm^{-1}$), mainly limited by the phase-matching condition in a GaSe crystal used for the intra-pulse difference frequency generation (IDFG) [9-11], while that of the Raman spectrum was about 3,000 $cm^{-1}$. One alternate solution for expanding the IR spectral bandwidth might be using a pump source in a longer wavelength (1.5-2.0 µm), where a broader IDFG generation can be possible under a relaxed phase-matching condition. However, it is technically challenging to develop a 10-fs-level pulsed laser with an average power of 100s mW [10, 12] (which is required for effective excitation of the IDFG and coherent Raman scattering) in the longer wavelength region, while such specifications are readily available in the 800 nm region with a commercially available Ti:S laser. Also, coherent Raman scattering spectroscopy with the longer-wavelength lasers has a disadvantage in biomedical applications, such as larger water absorptions and lower beam focusability. Therefore, expanding the spectral bandwidth of CVS with the 10-fs laser at 800 nm is a promising way.

In this study, we demonstrate ultra-broadband CVS with a 10-fs Ti:S mode-locked laser by implementing a cascaded-IDFG process with two nonlinear crystals (GaSe and $LiIO_3$), which vastly expands the IR spectrum. We show the CVS spectral bandwidth broader than 2000 $cm^{-1}$, more than double the previous demonstration. As a proof-of-principle demonstration, we measure complementary vibrational spectra of liquid-phase molecules covering the fingerprint and high-wavenumber C-H stretching regions.

Figure 1 shows the schematic of the ultra-broadband CVS. A light source is a 10-fs Ti:S mode-locked laser running at a repetition rate of 75 MHz, covering a broadband NIR spectrum from 700 to 910 nm. A NIR pulse from the laser splits into two in a Michelson interferometer consisting of a polarizing beamsplitter and quarter-wave plates. The delay between the double NIR pulses can be scanned over 2 cm at a rate of around 0.8 Hz with a free-space mechanical delay line. The pulses are temporally compressed by a chirped-mirror pair (CMP) after passing through an optical long-pass filter (LPF) with a cut-off wavelength of 700 nm. The pulses with an average power of 560-610 mW are focused onto a $LiIO_3$ crystal (United Crystals) with a thickness of 100-200 µm (The ambiguity of the thickness specified by the manufacturer) by using an f=25.4 mm off-axis parabolic mirror (OAPM). The polarization of the incident pulses is adjusted with a half-wave plate (HWP). The IDFG process in the $LiIO_3$ crystal generates a pair of MIR pulses. Another OAPM collects the generated MIR pulses together with the undepleted NIR pulses. The NIR pulses are spatially separated from the MIR pulses with a dichroic mirror (DM) and recompressed with another CMP. Then, the NIR pulses with an average power of 360-380 mW are focused onto a 30-µm-thick GaSe crystal (Eksma Optics) with an f=25.4 mm OAPM for the second stage IDFG. An LPF with a cut-off wavelength of 6 µm (Edmund, #36-151) spatially and spectrally combines the two pairs of MIR pulses generated by the cascaded IDFG process. The final undepleted NIR pulses are again compressed with a CMP and combined with the MIR pulses using a DM. An OAPM with a focal length of 15-mm or 12.7-mm focuses all the NIR and MIR pulses onto a sample kept in a homemade cuvette consisting of a pair of KBr windows with a Teflon spacer. Note that these pulse pairs are temporally separated to avoid unnecessary nonlinear optical phenomena at the sample position.

A MIR and a NIR photoreceiver detect the spectrally- and spatially-separated MIR and NIR pulses, respectively. The MIR detector, an $N_2$-cooled HgCdTe (MCT) detector (KLD-0.5-J1-3/11, Kolmar Technologies), measures interferograms in the temporal domain, and the signals are electrically low-pass-filtered and recorded with a digitizer (ATS9440, AlazarTech). The MIR average power at the detector is about 50 nW. Then, the fast Fourier transform (FFT) converts the interferograms to MIR spectra (CVS-IR spectra). On the other hand, a nonlinear FT-CARS process with the NIR pulses provides CVS-Raman spectra [8]. The first NIR pulse excites Raman-active molecular vibrations through the impulsive stimulated Raman scattering, and the second NIR pulse probes them. By scanning the delay between the pulses, the second pulses acquire periodic spectral shifts due to the refractive index modulations of the sample that arise from the molecular vibrations. The spectral shifts convert to intensity modulations of the pulses by filtering out the blue-shifted part of the spectra using short-pass filters (SPFs) with a cut-off wavelength of 700 nm. An avalanche photodetector (APD) (APD410A2/M, Thorlabs) measures the intensity modulations, i.e., the CARS interferograms. Finally, the FFT of the temporal waveform provides CARS spectra. We notice that the digitizer measures the IR and CARS interferograms simultaneously with two channels. Another digitizer channel measures reference continuous-wave interferograms generated with a HeNe laser, which are used for the wavenumber calibration of the CVS-IR and -Raman spectra. The calibration scheme based on the same reference signal provides precise wavenumber axes.

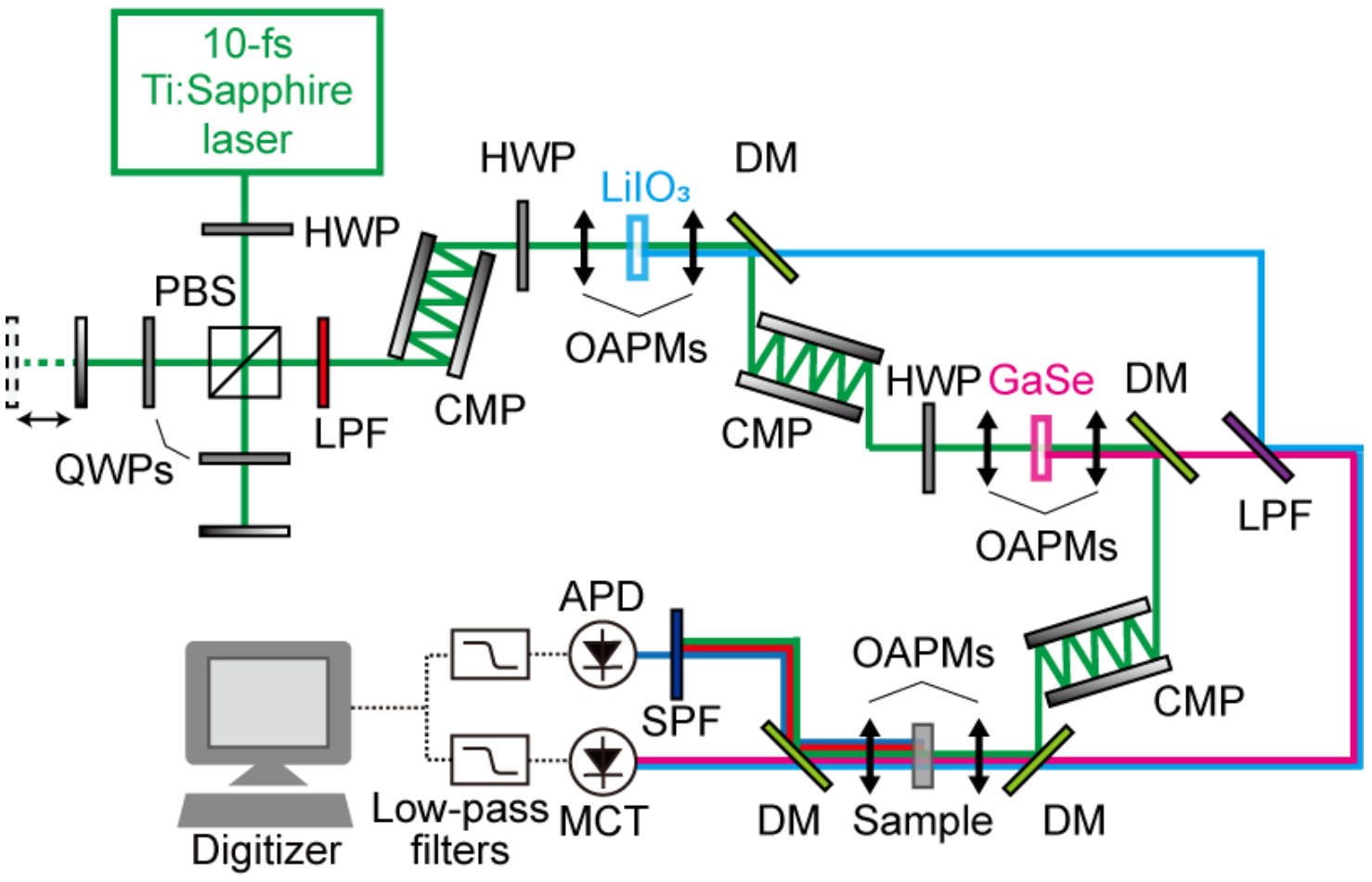

**Fig. 1.** Schematic of ultra-broadband CVS with cascaded IDFG. HWP: Half-wave plate, PBS: Polarizing beam splitter, QWP: Quarter-wave plate, LPF: Long-pass filter, CMP: Chirped-mirror pair, OAPM: Off-axis parabolic mirror, DM: Dichroic mirror, SPF: Short-pass filter, APD: Avalanche photodetector, MCT: HgCdTe detector.

We first characterize IR spectra generated via the cascaded IDFG process to diagnose the system. The 20-times averaged IR interferogram measured with the CVS system is shown in Fig. 2 (a). Since the CVS system has the interferometer before the IDFG processes, the measured raw temporal waveform contains additional signals on top of the linear interferogram. Due to the temporal overlap of the two NIR pulses at the nonlinear crystal, the slowly-varying intensity autocorrelation (IAC) and high-frequency fringes appear around the zero-path difference (ZPD) [7]. We numerically suppress the IAC and the high-frequency components by subtracting the fitted IAC and low-pass-filtering, respectively. The IR spectrum FFTed from the interferogram is shown at the bottom of Fig. 2 (b) (colored in blue-green), spanning from 800 to 2900 cm$^{-1}$. The cut-off wavelength of the MCT detector and the phase-matching condition of the $LiIO_3$ crystal determine the lowest and highest wavenumber edges, respectively. The top (red) and middle (blue) spectra in Fig. 2 (b) represent the MIR spectra generated from the GaSe and $LiIO_3$, respectively, which are individually measured by blocking the other's beam paths. The dips around 1,600 and 2,350 cm$^{-1}$ attribute to the ambient $H_2O$ (and cut-off of the 6-µm LPF) and $CO_2$ absorptions, respectively.

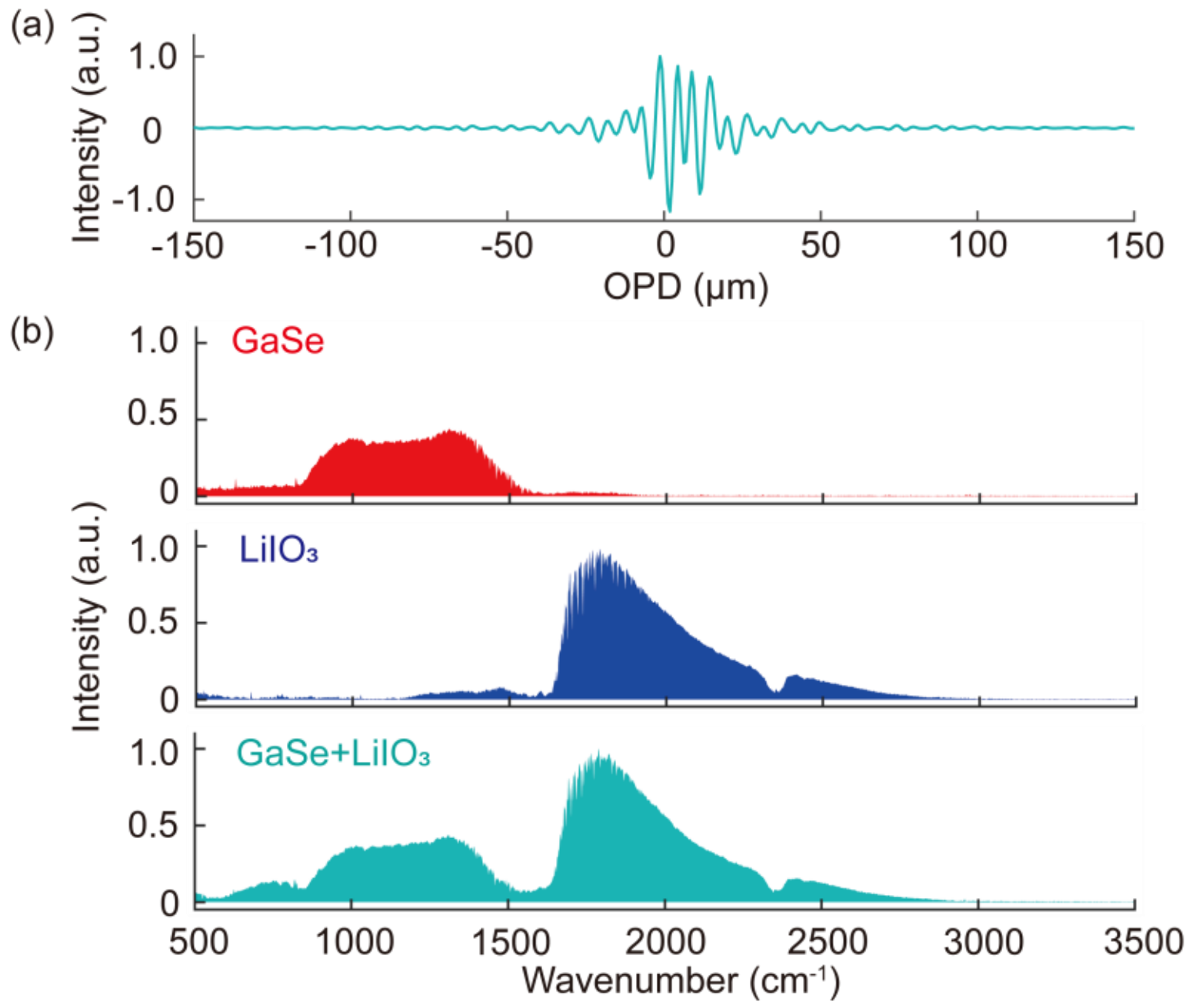

**Fig. 2.** Broadband IR spectrum generated via cascaded IDFG. (a) IR interferogram measured with the ultra-broadband CVS system. (b) The bottom panel shows the FFT spectrum of the IR interferogram (blue-green). The top (red) and middle (blue) panels show the individually measured IR spectra generated from the GaSe and the $LiIO_3$, respectively.

Next, we measure liquid phenylacetylene with the CVS system to show simultaneous ultra-broadband IR and Raman spectroscopy capability. Figure 3 shows the measured IR and Raman interferograms. The synchronously appeared center-bursts of the interferograms verify the simultaneous measurement. The inset in Fig. 3 shows enlarged views of the interferograms. The IR and Raman molecular vibrations of the liquid phenylacetylene are clearly observed.

Figure 4 (a) shows a broadband IR transmittance spectrum measured with the CVS spectrometer compared with a reference spectrum measured with a commercial FTIR spectrometer (FTIR-6800, JASCO). The CVS-IR spectrum is obtained by Fourier-transforming the coherently averaged 30 interferograms with triangular apodization. The transmittance spectra measured with an unapodized spectral resolution of 9 cm$^{-1}$ are retrieved from spectra with and without the sample. The CVS spectrometer is enclosed with an $N_2$ purge box to suppress the ambient gaseous absorption. We confirm that the CVS-IR spectrum agrees well with the reference spectrum with an assignment of the several

IR absorption bands of phenylacetylene: C-H bendings at 1,026 and 1072 $cm^{-1}$, a broad overtone and combination at 1235 $cm^{-1}$, ring stretchings at 1442 and 1484 $cm^{-1}$, aromatic combinations at 1677, 1758, 1809, 1886, and 1955 $cm^{-1}$, and a C≡C stretching at 2110 $cm^{-1}$ [13]. The signal-to-noise ratio (SNR) of a single-shot non-averaged spectrum is ~150, which is evaluated from the peak intensity of the single-shot spectrum and the standard deviation of the noise at the region outside of the IR spectrum.

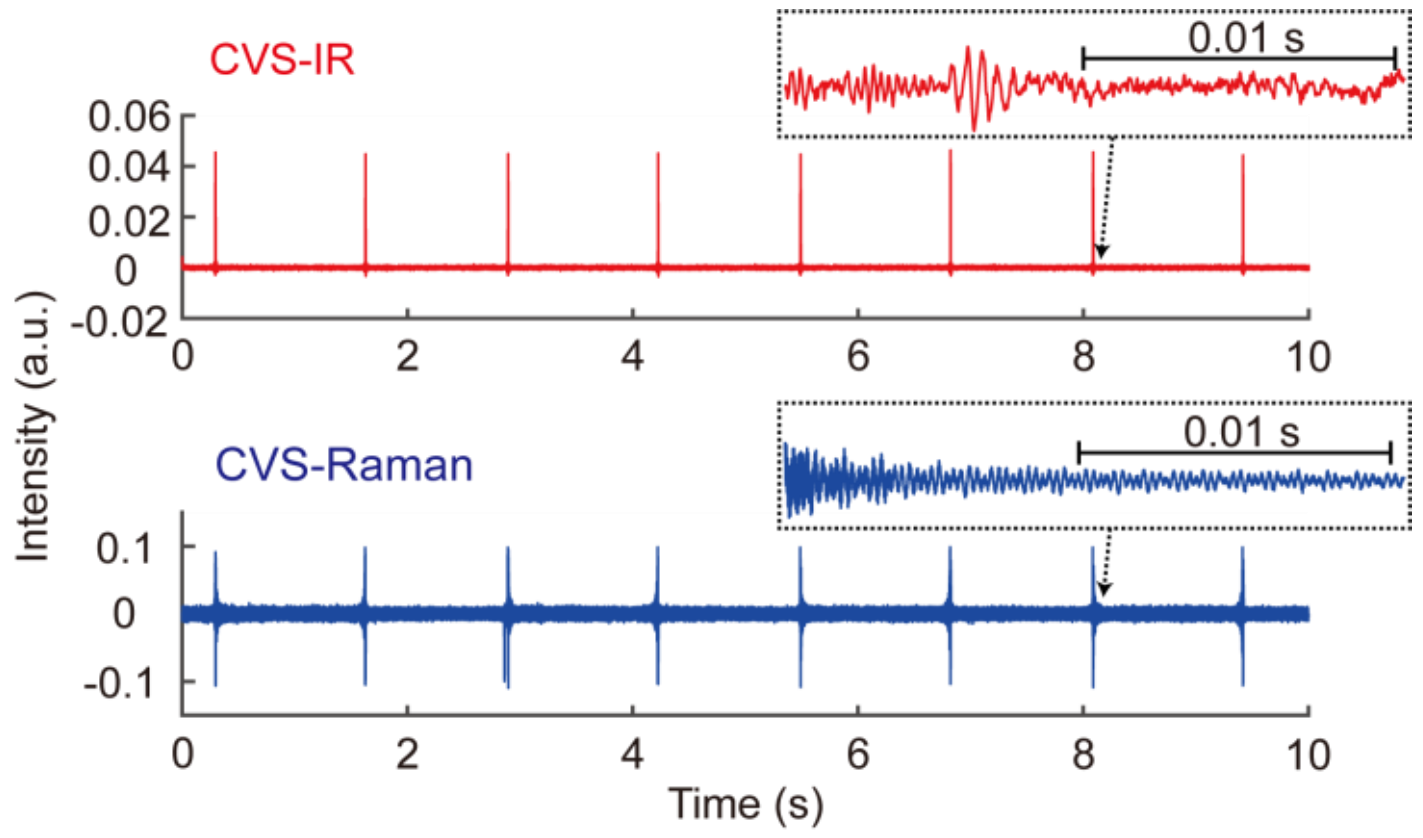

**Fig. 3.** Sequential IR (red) and CARS (blue) interferograms measured with the ultra-broadband CVS. The inset shows the enlarged view of the IR and Raman molecular vibrations.

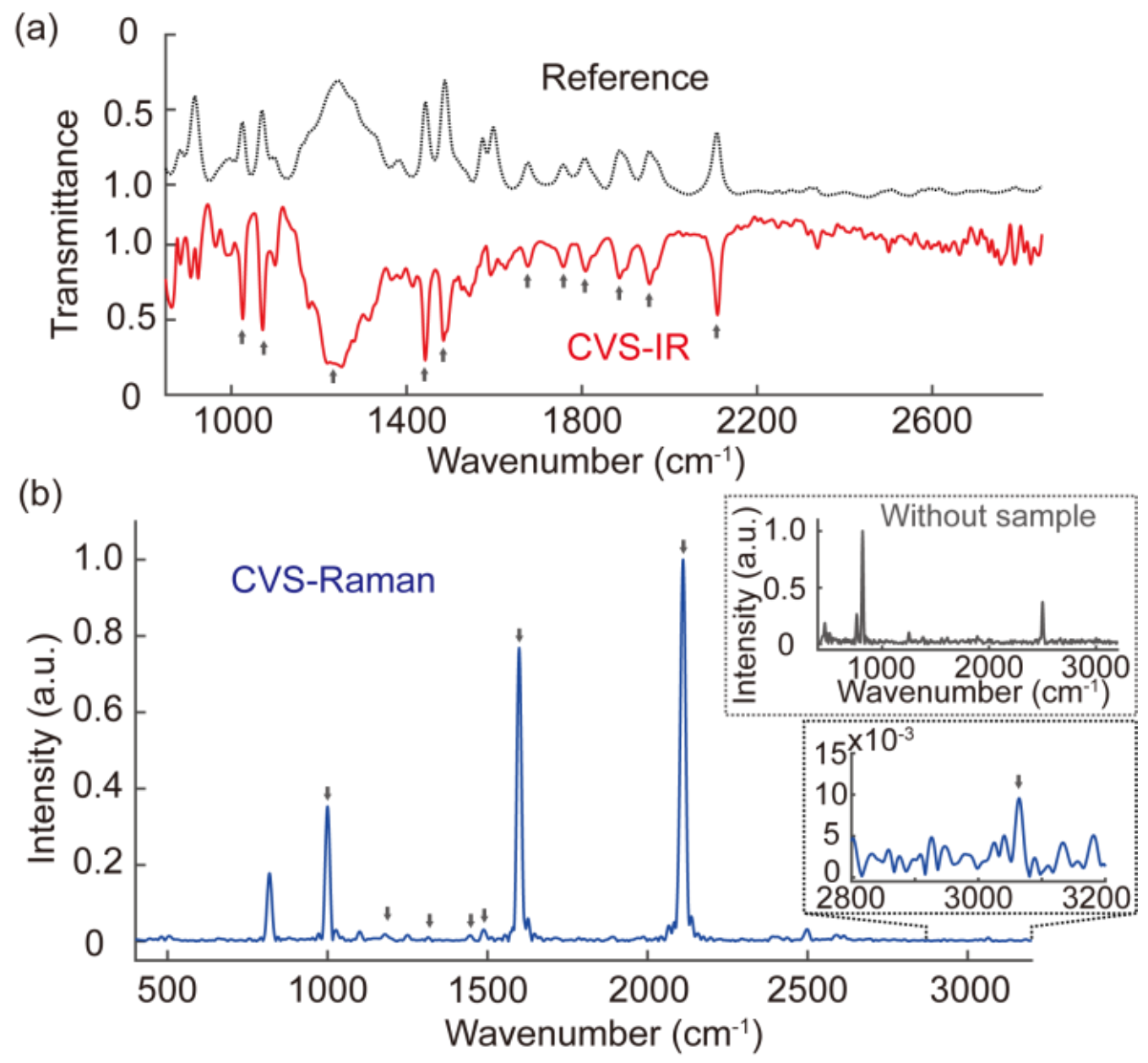

**Fig. 4.** Complementary vibrational spectra of phenylacetylene (a) IR spectra of liquid phenylacetylene spanning from 850-2850 $cm^{-1}$ measured with the ultra-broadband CVS (solid red) and a standard IR spectrometer (dotted black). The small arrows represent the assigned IR spectral bands. (b) Raman spectrum of phenylacetylene (blue). The upper inset shows a normalized spectrum obtained without a sample. The lower inset shows the enlarged

view of the phenylacetylene Raman spectrum around 3000 cm$^{-1}$. The small arrows represent the assigned Raman spectral bands.

Figure 4 (b) shows an FFT result of a 30-averaged CARS interferogram with triangular apodization, whose unadpozied spectral resolution is 9 cm$^{-1}$. The large non-resonant signals in the interferograms around ZPD are omitted from the FFT window to avoid spectral distortion. We assign several Raman bands in the CVS-Raman spectrum: a ring breathing at 1000 cm$^{-1}$, a C-H bending at 1179 cm$^{-1}$, an overtone at 1315 cm$^{-1}$, ring stretchings at 1446, 1488, and 1599 cm$^{-1}$, a C≡C stretching at 2111 cm$^{-1}$, and a C-H stretching at 3064 cm$^{-1}$ [13]. As shown in the lower inset of Fig. 4(b), the highest observable wavenumber of the CVS-Raman spectrum is ~3100 cm$^{-1}$. The cut-off wavelengths of the LPF and SPF determine the lowest observable wavenumber of ~100 cm$^{-1}$. Raman peaks around 1200 cm$^{-1}$ look weaker than the literature, probably due to the spectrally-non-uniform excitation efficiency caused by the slightly distorted temporal pulse shape caused by the higher-order dispersion, which is not compensated in our experiment. There are some peaks not originating from phenylacetylene, which can also be seen in a spectrum without the sample (Upper inset of Fig. 4(b)). We attribute that the two peaks at 700-850 cm$^{-1}$ originate from the CARS signal generated from the $LiIO_3$ crystal [14], while the other spikes around 500 cm$^{-1}$, 1250 cm$^{-1,}$ and 2500 cm$^{-1}$ from instrumental noises. Since the CARS signals from the $LiIO_3$ crystal are strong, the NIR beams at the $LiIO_3$ crystal are slightly defocused so that they become weaker than those from the samples. A single-shot SNR of the C≡C stretching band is ~200 when the average power of the NIR excitation pulses is ~180 mW at the sample.

Finally, we measure complementary vibrational spectra of other liquid samples: benzene, chloroform, and toluene. Figure 5 shows 30-times averaged spectra with an unapodized resolution of 9 cm$^{-1}$. The noise spikes in the CVS-Raman spectra are eliminated by subtracting the background spectrum without the samples. IR and Raman bands in the measured spectra are assigned as follows. In the IR spectrum of benzene, it is observed a C-H bending at 1039 cm$^{-1}$, a ring stretching+deformation at 1479 cm$^{-1}$, and aromatic combinations at 1816 and 1961 cm$^{-1}$ [15, 16]. In the Raman spectrum, it is observed a ring breathing at 993 cm$^{-1}$, a ring stretching at 1586 cm$^{-1}$, and a C-H stretching at 3063 cm$^{-1}$ [15]. In the IR and Raman spectra of chloroform, it is observed a C-H bending at 1215 cm$^{-1}$ and a C-Cl stretching at 670 cm$^{-1}$, respectively [15]. The small peaks at 2402 cm$^{-1}$ in the IR spectrum might be overtone bands of the C-H bending modes. In the IR spectrum of toluene, it is observed C-H bendings at 1178 and 1500 cm$^{-1}$, ring stretchings at 1030, 1081, and 1605 cm$^{-1}$, $CH_3$ deformations at 1380 and 1456 cm$^{-1}$, and aromatic combinations at 1804, 1858, and 1943 cm$^{-1}$ [17, 18], while in the Raman spectrum, ring-$CH_3$ stretchings at 787 and 1209 cm$^{-1}$, a ring-breathing at 1004 cm$^{-1}$, ring-stretchings at 1030 and 1605 cm$^{-1}$, a $CH_3$ deformation at 1381 cm$^{-1}$, and C-H stretchings at 2917 and 3054 cm$^{-1}$ [17, 18].

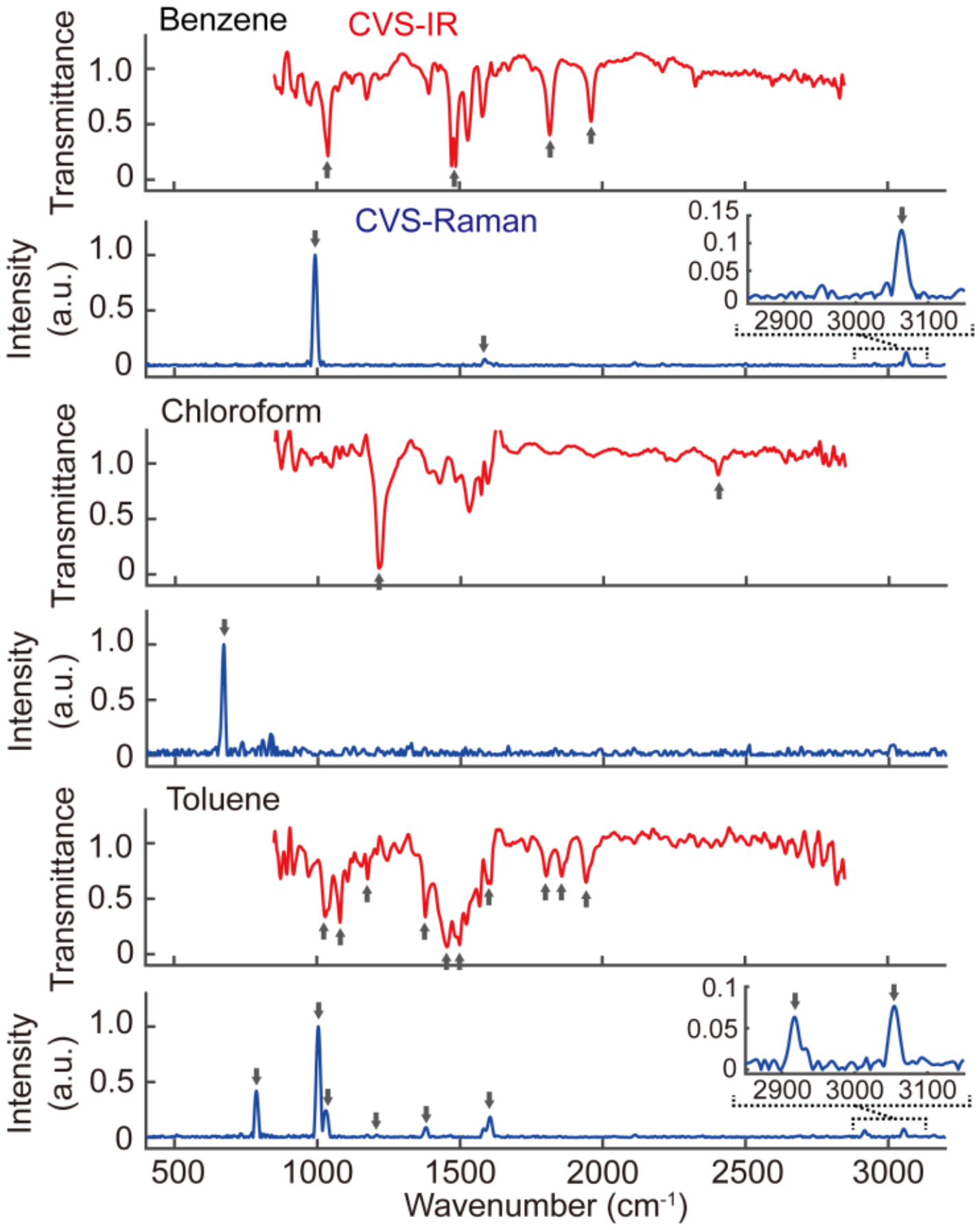

**Fig. 5.** Complementary vibrational spectra of liquid samples. The upper, middle, and bottom panels show the spectra of benzene, chloroform, and toluene, respectively. Red and blue correspond to the IR and Raman spectra, respectively. The insets show the enlarged views of the Raman bands of benzene and toluene that appear around 3000 $cm^{-1}$. The small arrows represent the assigned IR and Raman spectral bands.

In conclusion, we demonstrated the ultra-broadband CVS system with the cascaded IDFG, which can measure IR and Raman spectra spanning over 2000 $cm^{-1}$. The system can be improved further by several modifications. The sensitivity can be enhanced by applying several methods, such as electro-optic sampling [9,19] or heterodyne detection [20]. The IR spectral range can be broadened by adding other nonlinear crystals into the system. The spectral acquisition rate can be increased by implementing high-scan-rate FTS techniques such as dual-comb spectroscopy [21], rapid-scan FTS [22], or phase-controlled FTS [23]. The CVS system can also be applied to microscopy [24], providing multimodal images. In addition to the hardware improvements, some numerical methods such as the two-dimensional correlation analysis [25] or the Matrix Pencil (MP) method [26] can enhance the capability of the system. Finally, the compressive sensing technique allows for interferometric random data sampling to retrieve sparse broadband vibrational spectra, reducing measurement data points [27].

### Funding
Japan Society for the Promotion of Science (20H00125); Precise Measurement Technology Promotion Foundation (PMTP-F)

### Acknowledgments
The authors thank Junji Yumoto and Makoto Gonokami for the use of their equipment.

### Disclosures
The authors declare no competing interests.

### Data availability
The data that support the findings of this study are available from the corresponding author upon reasonable request.